\documentclass[american]{article}
\usepackage[T1]{fontenc}
\usepackage{geometry}
\usepackage{units}
\usepackage{multicol}
\usepackage{amsmath}
\usepackage{amsthm}

\makeatletter
\theoremstyle{remark}
\newtheorem*{acknowledgement*}{\protect\acknowledgementname}

\makeatother

\usepackage{babel}
\usepackage[style=numeric,sorting=none]{biblatex}
\providecommand{\acknowledgementname}{Acknowledgement}

\begin{document}

\title{The rationality of irrationality in the Monty Hall problem}

\author{Torsten En{\ss}lin and Margret Westerkamp\\
{\small{}Max Planck Institute for Astrophysics, Karl-Schwarzschild-Str.
1, 85741 Garching, Germany}}
\maketitle
\begin{abstract}
The rational solution of the Monty Hall problem unsettles many people.
Most people, including the authors, think it feels wrong to switch
the initial choice of one of the three doors, despite having fully
accepted the mathematical proof for its superiority. Many people,
if given the choice to switch, think the chances are fifty-fifty between
their options, but still strongly prefer to stay with their initial
choice. Is there some sense behind these irrational feelings? 

We entertain the possibility that intuition solves the problem of
how to behave in a real game show, not in the abstract textbook version
of the Monty Hall problem. A real showmaster sometimes plays evil,
either to make the show more interesting, to save money, or because
he is in a bad mood. A moody showmaster erases any information advantage
the guest could extract by him opening other doors which drives the
chance of the car being behind the chosen door towards fifty percent.
Furthermore, the showmaster could try to read or manipulate the guest's
strategy to the guest's disadvantage. Given this, the preference to
stay with the initial choice turns out to be a very rational defense
strategy of the show's guest against the threat of being manipulated
by its host. Thus,  the intuitive feelings most people have about
the Monty Hall problem coincide with what would be a rational strategy
for a real-world game show. Although these investigations are mainly
intended to be an entertaining mathematical commentary on an information-theoretic
puzzle, they touch on interesting psychological questions.
\end{abstract}
\vspace{1cm}

\begin{multicols}{2}

\section*{The Monty Hall problem}

The textbook Monty Hall problem \cites{Monty-Hall}{Wikipedia} goes
as follows. A showmaster presents you three doors to choose from.
Behind one is a sports car that you win if you decide to open this
door. Behind the others are goats that have no value to you. You indicate
a choice, but before you open the chosen door, the showmaster opens
another one, revealing a goat, and gives you the opportunity to revise
your choice. Should you stay with your choice or should you switch
to the other door?

The implicit assumption is that the showmaster always behaves in this
way. In this case the optimal strategy can be found by various lines
of argument. Your initial probability of picking the right door is
$\nicefrac{1}{3}$. If your strategy is to always stay with your choice,
this is your probability of winning the car, irrespective of whether
the showmaster opens zero, one, or both other doors. In $\nicefrac{2}{3}$
of the cases the car is behind one of the other doors. The showmaster
is so friendly as to indicate which of those is certainly wrong. So,
if you switch, and the car is behind one of the other doors, he guides
you to the good one. This will work in $\nicefrac{2}{3}$ of the cases,
which is then the chance of winning if you switch.

So far so good. Logic has shown us what to do best. But -- it just
feels wrong. It feels so wrong that approximately $10,000$ readers
of Marilyn vos Savant's column ``Ask Marilyn'' write to her to say
that she must be erring after she had explained the correct solution
to the Monty Hall problem. Among them, nearly $1,000$ holders of
PhDs \cite{Times}. Why don't they (and we) trust Marilyn? Well, probably,
because for this we also have to trust the showmaster! 

\section*{The evil showmaster}

Imagine the textbook assumption is wrong that the showmaster always
behaves in the described way. Real showmasters can and occasionally
do behave differently. Imagine the worst case: the showmaster is \emph{evil}
\cite{Evil-Monty} and opens other doors only if you pick the door
with the car. Thus, if you pick a door with a goat, he will immediately
open it and you lose. This happens in $\nicefrac{2}{3}$ of the cases.
In $\nicefrac{1}{3}$ of the cases you pick the door with the car.
Then, the evil showmaster opens another one in the hope that you play
switch as you have learned this to be superior. In that case you lose
as well. So if you always play switch and the showmaster plays evil,
you always lose. However, if you play stay, you win in $\nicefrac{1}{3}$
of the cases, namely, whenever you initially picked the door with
the car. The showmaster cannot change this. Just be stubborn and ignore
the seduction of a virtual additional chance of $\nicefrac{1}{3}$
by switching and you defend the guaranteed $\nicefrac{1}{3}$ winning
chance of your initial choice.

A showmaster always playing evil would save the broadcasting company
real money. Only $\nicefrac{1}{3}$ of the stay-people get cars, and
all the switch-people get only goats. However, he might ruin the show.
It would become boring and people would stop watching it. Therefore,
he should behave in an evil way only with some frequency $p$. And
he should tailor this frequency to create maximal suspense while cutting
financial losses.

\section*{Optimal moodiness}

In order to optimally choose his frequency $p$ of playing evil the
showmaster has to consider the situation of his guests. For this,
he had better assume that whatever frequency $p$ he chooses, this
will also be known to his guests. They might obtain this number by
studying broadcasts of the show or by performing the same calculation
as he does to determine his optimal moodiness, as we will do in the
following as well. 

Imagine that you have made your initial choice, and he has opened
another door, revealing a goat. Should you now stay or switch? In
the fraction $p$ of the cases when his mood is evil you had better
stay, ensuring your winning probability of $\nicefrac{1}{3}$. If
you switch in this case, you surely lose. In the fraction $1-p$ of
the cases, when he plays the fair Monty Hall textbook showmaster,
you win in $\nicefrac{1}{3}$ of the cases if you stay and in $\nicefrac{2}{3}$
if you switch. Multiplying and adding these cases together yields
that you win with playing staying in $p\times\nicefrac{1}{3}+(1-p)\times\nicefrac{1}{3}=\nicefrac{1}{3}$
of all cases. If you play switching, you win in $p\times0+(1-p)\times\nicefrac{2}{3}=(1-p)\times\nicefrac{2}{3}$
of the cases. Which strategy is better depends on $p$. If $p<\nicefrac{1}{2}$,
switching is better, if $p>\nicefrac{1}{2}$, staying is better, and
for $p=\nicefrac{1}{2}$ the strategy does not matter, you win in
$\nicefrac{1}{3}$ of all cases. 

The showmaster should make you face a fifty-fifty chance of him being
evil, as this will put you under maximal stress. This is what the
audience wants to see and therefore this is what the showmaster will
probably aim for. He might add or subtract a small margin, in order
to save a bit money or to make the show more attractive, respectively,
but this is hard for you to judge. He will make sure that you are
nearly clueless about his mood by choosing $p\approx\nicefrac{1}{2}$.
This optimal amount of evilness is therefore what you better assume.
He will be moody just to the level that erases any information advantage
you could have gotten from having observed his action.

\section*{The information game}

The probability $p=P(\mbox{evil})$ describes your belief before the
game that the showmaster has evil intentions. During the game the
probability you assign to him being evil will change. 

If he opens the door that you initially picked and reveals a goat
there (``my'' for ``he opens my door''), then you can be sure
that he is evil, $P(\text{evil | my})=1$, as the ``fair'' showmaster
never does this, but always opens an ``other'' door, $P(\text{my | fair})=1-P(\text{other | fair})=0$.
We assume here that no showmaster ever reveals the sports car without
you insisting on choosing its door, $P(\text{other | car})=1$, where
with ``car'' and ``goat'' we label what is behind your initially
chosen door. 

In the other case, where he opens another door, the situation is less
clear. However, since some of the evil possibilities are ruled out,
your confidence in him being in a fair mood has increased. How much
can be worked out using Bayes' theorem as we have our prior beliefs
\[
P(\text{evil)}=p\mbox{ and }P(\text{fair})=1-p
\]
and can specify the likelihoods of the different events:
\begin{align*}
P(\text{other | fair}) & =1, & P(\text{my | fair)} & =0,\\
P(\text{other | evil}) & =\frac{1}{3}, & P(\text{my | evil}) & =\frac{2}{3}.
\end{align*}

Going through the maths of Bayes' theorem we find 
\begin{eqnarray*}
P(\mbox{evil }|\mbox{ other}) & = & \frac{P(\mbox{other, evil})}{P(\text{other)}}\\
 & = & \frac{P(\mbox{other }|\mbox{ evil})\,P(\text{evil})}{P(\text{other, evil})+P(\text{other, fair})}\\
 & = & \frac{\frac{1}{3}\times p}{\frac{1}{3}\times p+1\times(1-p)}\\
 & = & \frac{p}{3-2\,p}.
\end{eqnarray*}
We have $P(\mbox{evil }|\mbox{ other})=\nicefrac{1}{4}$ for $p=\nicefrac{1}{2}$.
Thus, having seen the showmaster opening another door makes you more
confident in him being fair. Should you therefore trust him and open
the remaining door as now your probability of him being evil is well
below $\nicefrac{1}{2}$?

Not at all! You should still play carefully, as the threshold of $p=\nicefrac{1}{2}$
was derived for the prior probability $P(\mbox{evil})$ and not for
the posterior probability $P(\mbox{evil }|\mbox{ other})$. In fact
all that matters to you is $P(\mbox{car |\mbox{ other})}$, the probability
that the car is behind your initially picked door, irrespective of
whether the showmaster is evil or not. This requires you to marginalize
out the momentary mood of the showmaster, but not his action, as the
latter is important information. For calculating this chance, we recall
that $P(\text{other | car})=1$ as any showmaster, regardless of fair
or evil, will open another door if you picked the car. If you picked
a goat, only the fair one will open another door. This will therefore
happen with probability $P(\text{other | goat})=1-p$. The probabilities
of seeing another door opened given what is behind your door now permit
us to work out 

\begin{eqnarray*}
P(\mbox{car |\mbox{ other})} & = & \frac{P(\mbox{other, car})}{P(\mbox{other)}}\\
 & = & \frac{P(\text{other | car})\,P(\text{car})}{P(\mbox{other, car})+P(\mbox{other, goat})}\\
 & = & \frac{1\times\frac{1}{3}}{1\times\frac{1}{3}+(1-p)\times\frac{2}{3}}\\
 & = & \frac{1}{3-2p}.
\end{eqnarray*}
Thus, a showmaster with evil intentions half of the time leaves us
clueless behind which of the remaining doors the car is to be expected,
since $P(\mbox{car |\mbox{ other})}=\nicefrac{1}{2}$ for $p=\nicefrac{1}{2}$.
This value is the intuitive feeling of many people about their chance,
indicating that they also anticipate a fifty-fifty percent chance
of the showmaster playing evil, an option a billion years of evolution
have taught us never to forget. Or at least, our feeling tells us
that the showmaster is manipulating us and probably has erased any
information we could have obtained on the correct door.

In contrast to this, a showmaster always exhibiting the same mood
would provide you with information on the door with the car. His message,
however, would depend on his mood. A showmaster always playing fairly
($p=0$) would leave only a winning chance of $P(\mbox{car |\mbox{ other, fair})}=\nicefrac{1}{3}$
for your door, but $\nicefrac{2}{3}$ for the remaining door. The
always evil showmaster ($p=1$) would inform us by opening another
door that we picked the right one, as $P(\mbox{car |\mbox{ other, evil})}=1$
then. To benefit from this message, you would need to have identified
his mood. But beware, he might have read your mind as well.

\section*{The mind-reading showmaster}

If a fraction $q=P(\text{stay})$ of the guests of the show stay with
their doors and $1-q=P(\text{switch})$ of the guests switch to the
other door, the fraction of guests winning the car is
\begin{eqnarray*}
P(\text{win}) & = & P(\text{stay)}\,P(\text{car})+P(\text{switch})\,P(\text{fair})\,P(\text{goat})\\
 & = & q\times\frac{1}{3}+(1-q)\times(1-p)\,\times\frac{2}{3}\\
 & = & \frac{2-2\,p-q+2\,p\,q}{3}
\end{eqnarray*}
as ``stay'' wins irrespective of the showmaster's strategy if the
car is behind the initial door, and \textquotedbl switch\textquotedbl{}
wins if there was a goat and the showmaster plays fair. For the optimally
moody showmaster with $p=\nicefrac{1}{2}$ this means $P(\text{win})=\nicefrac{1}{3}$
as the guest strategy then does not matter. The always fair showmaster
is significantly more expensive for the broadcasting company, as for
$p=0$ we have $P(\text{win})=\nicefrac{(2-q)}{3}\ge\nicefrac{1}{3}$
and thus the switching guests double their chance. The always evil
showmaster ($p=1$) saves real money with $P(\text{win})=\nicefrac{q}{3}\le\nicefrac{1}{3}$
as all switching guests lose surely. However, he can follow this strategy
only until the reputation of the show is ruined.

Anyhow, such an economic winning rate could also be achieved by a
mind-reading showmaster, who adapts his strategy to the situation.
He could play fair whenever there is a guest expected to play stay
or one who will switch to a goat. Otherwise he plays evil. Then the
number of winning guests would be that of the always evil master.
But the showmaster would look much more generous than the evil one,
as he opens other doors more frequently. 

If the showmaster can read his guest's mind perfectly, the winning
rate is $P(\text{win})=\nicefrac{q}{3}\le\nicefrac{1}{3}$, but he
opens other doors in 
\begin{eqnarray*}
P(\text{other | mind-reader}) & = & P(\text{stay})+P(\text{switch})\,P(\text{car})\\
 & = & q+(1-q)\times\frac{1}{3}\\
 & = & \frac{1+2\,q}{3}
\end{eqnarray*}
of the games, more often than the $\nicefrac{1}{3}$ if always playing
evil. This will look generous, given that most people prefer to stay
in fear of an evil showmaster, meaning $q>\nicefrac{1}{2}$, which
implies $P(\text{other | mind reader})>\nicefrac{2}{3}$. With a bit
of mind-reading, the showmaster can keep a reputation of being very
fair, while in fact saving real money for his company. And in $\nicefrac{2}{3}$
of the games with guests that play stay, it will appear in the end
that switching would have been the better choice. This should motivate
enough of the people studying the show to use the switching strategy
when they happen to play. As those switchers will never win, they
contribute to the benefit of the company. 

Is this possibility realistic? Well, a good showmaster worth his money
should be able to read his guest's mind. Before the game starts he
interacted enough to get a good guess on the preferred strategy of
the guest. The showmaster might even manage to manipulate the guest
into adopting a certain strategy. Using this against his guest can
make him look more fair than he is in reality. And if the showmaster
is unsure, he just plays evil, to be on the safe side.

\section*{The acting guest}

There is also a counter strategy available to the guest against the
mind-reading showmaster. If the guest manages to act in a way that
the showmaster believes the guest is playing stay, whereas the guest
actually will play switch, the winning chance of the guest becomes
$\nicefrac{2}{3}$. This, however, won't work too often, as the showmaster
will certainly recognize whenever he is being fooled. If this happens
too often, he might return to the randomized strategy. Or, if he learns
to identify the acting guests, he just plays evil on them and they
go home with only a goat. Thus the acting guest takes a real risk.
If caught, he loses for sure.

\section*{}

\section*{}

\section*{The real show}

Most people will not perform the above mathematical consideration
explicitly when faced with the situation of a real game show. Their
intuition, however, might take many factors into account that the
classical textbook version of the Monty Hall problem ignores. Intuition
knows that the showmaster could have acted differently, might be evil,
that he wants to create a situation of emotional tension, and probably
wants to avoid giving away too many expensive cars. The showmaster
will likely be perceived as an experienced veteran of many such shows,
being well capable of guessing or manipulating the strategy of individual
guests. Even if told that the showmaster will always open another
door, intuition won't put too much belief in this claim. There is
just too much circumstantial evidence coded into our intuition of
real showmasters, and other predators, to behave differently, usually. 

As the Monty Hall show is an information game, it would be very natural
for the showmaster to try to erase any information his guest has on
the proper door. The existence of such an information-erasing strategy
for the showmaster might determine the ordinary guest's intuition.
If it is possible for the show master to erase any knowledge, he probably
does so and therefore a $\nicefrac{1}{2}$ chance to the car being
behind their door is better assumed.

It should be said that psychology offers also another very plausible
explanation for the assignment of a $\nicefrac{1}{2}$ chance. It
is well known that people use logical shortcuts in their reasoning,
which should also lead to the same result. In psychological investigations
of the Monty Hall problem the ``insensitivity to prior probability
outcomes'', considered by \textcite{Tversky&Kahneman}, is often
mentioned as an explanation of the guest's behavior. It states that
prior probabilities are ignored by humans when given new evidence,
even if the new information is worthless. Transferring this to the
Monty Hall problem, one can explain why people assign a $\nicefrac{1}{2}$
probability of winning for both strategies, switching or staying.
Thus, it could well be a mere coincidence that people's intuition
provides here a good guess for the chances. 

Still, the preference that people show for staying is not explained
by this. Staying is better in case the show master is more evil, mind-reading,
or manipulative, as it guarantees a winning chance of $\nicefrac{1}{3}$,
whereas switching risks to always loose. The only safe strategy against
the maneuvers of the showmaster is to stay with the initial choice,
thereby ensuring that one will bring home a car in $\nicefrac{1}{3}$
of the cases. 

The preference for staying might therefore be explained by the distrust
in the intentions of the show master. The opening of the second door
is easily perceived as an intentional reaction of the show master
to the guest's move to pick a door. Psychological studies by \textcite{Chance&Deception}
showed that people pay special attention to whether effects are caused
by humans or non-humans. Humans have intentions and these can be bad.
As the human brain is evolutionary trained to watch out for deception,
the more risky strategy to switch is disfavored, despite the $\nicefrac{1}{2}$
chance assignment people make. 

Thus, the apparently irrational strategy of staying turns out to be
in fact very rational, as it is immune against all the mean tricks
the showmaster might use. And the apparent contradiction between people
assigning a $\nicefrac{1}{2}$ chance -- for whatever psychological
reason -- to the car being behind the other door, but strongly insisting
on staying with their chosen door is also lifted. The former is an
estimate with some uncertainty attached to it. The latter is a decision,
taking potential losses into account, which tells us that we could
be a sure loser if the showmaster is mentally above us. 

Did the real Monty Hall knew about all this? We can let him answer
this by himself \cite{Gruber2010}: \textquotedbl But if he {[}the
showmaster{]} has the choice whether to allow a switch or not, beware.
Caveat emptor. It all depends on his mood.\textquotedblright{} 
\begin{acknowledgement*}
We thank Karen Redinger Emmendorfer, Tim Sullivan, Martin Reinecke,
Ancla Müller, Sebastian Hutschenreuter, Reimar Leike, and two anonymous
reviewers for insightful comments and language corrections. 
\end{acknowledgement*}
\printbibliography
\end{multicols}

\end{document}